\documentstyle[12pt,pictex,epsfig,amssymb,amsbsy]{article}

\renewcommand{\thefootnote}{\fnsymbol{footnote}}

\newcommand{\bbox}[1]{\boldsymbol{#1}}
\newcommand {\defeq}{\stackrel{\rm def}{=}}
\newcommand{\tr}[1]{\:{\rm tr}\,#1}

\def\e{{\,\rm e}\,}

\def\d{{\rm d}}

\def\i{{\rm i}}

\newcommand{\rf}[1]{(\ref{#1})}
\newcommand{\eq}[1]{Eq.~(\ref{#1})}
\def\be{\begin{equation}}
\def\ee{\end{equation}}
\def\beq{\begin{equation}}
\def\eeq{\end{equation}}
\def\bea{\begin{eqnarray}}
\def\eea{\end{eqnarray}}
\def\LA{\left\langle}
\def\RA{\right\rangle}
\def\N{{N_c}}
\def\A{{\cal A}}

\def\D{{\cal D}}
\def\G{{\cal G}}
\def\j{{\bbox j}}
\def\J{{\bbox J}}

\newcommand{\ie}{{\it i.e.}\ }

\newcommand{\ra}{\rightarrow}
\newcommand{\pintxx}{\int_x^x \hspace{-1.65em}\not\hspace{1.13em}}
\newcommand{\pint}{\int\hspace{-1.195em}\not\hspace{0.8em}}

\font\thinlinefont=cmr5
\begingroup\makeatletter\ifx\SetFigFont\undefined
\def\x#1#2#3#4#5#6#7\relax{\def\x{#1#2#3#4#5#6}}%
\expandafter\x\fmtname xxxxxx\relax \def\y{splain}%
\ifx\x\y   
\gdef\SetFigFont#1#2#3{%
  \ifnum #1<17\tiny\else \ifnum #1<20\small\else
  \ifnum #1<24\normalsize\else \ifnum #1<29\large\else
  \ifnum #1<34\Large\else \ifnum #1<41\LARGE\else
     \huge\fi\fi\fi\fi\fi\fi
  \csname #3\endcsname}%
\else
\gdef\SetFigFont#1#2#3{\begingroup
  \count@#1\relax \ifnum 25<\count@\count@25\fi
  \def\x{\endgroup\@setsize\SetFigFont{#2pt}}%
  \expandafter\x
    \csname \romannumeral\the\count@ pt\expandafter\endcsname
    \csname @\romannumeral\the\count@ pt\endcsname
  \csname #3\endcsname}%
\fi
\fi\endgroup

\begin{document}

\begin{titlepage}
\begin{flushright}
ITEP--TH--25/04\\
hep-th/0407028\\
July, 2004
\end{flushright}
\vspace{1.3cm}

\begin{center}
{\LARGE The First Thirty Years of Large-$N$ Gauge Theory} 
\\[.4cm]
\vspace{1.4cm}
{\large Yuri Makeenko}\footnote{E--mail:
makeenko@itep.ru \ \ \ \
 makeenko@nbi.dk \ } \\
\vskip 0.2 cm
{\it Institute of Theoretical and Experimental Physics,}
\\ {\it B. Cheremushkinskaya 25, 117259 Moscow, Russia}
\\ \vskip .1 cm
and  \\  \vskip .1 cm
{\it The Niels Bohr Institute,} \\
{\it Blegdamsvej 17, 2100 Copenhagen {\O}, Denmark}
\end{center}
\vskip 1.5 cm
\begin{abstract}
I review some developments in the large-$N$ gauge theory since 1974.
The main attention is payed to: multicolor QCD, matrix models,
loop equations, reduced models, 2D quantum gravity, free random variables,
noncommutative theories, AdS/CFT correspondence.

This talk was not given at the Workshop on Large-$N_c$ QCD, Trento, Italy,
July 5--10, 2004 because of the notorious ``visa problem''.

\end{abstract}

\end{titlepage}
\setcounter{page}{2}
\renewcommand{\thefootnote}{\arabic{footnote}}
\setcounter{footnote}{0}

\section*{Preface}

Large-$N$ gauge theories are with us since the work 
by 't~Hooft of 1974~\cite{Hoo74}.%
\footnote{Here and below I quote the year of a journal publication,
as it was custom in the pre-arXiv epoch.}
In this talk I review some milestones of their developments
since then. I pay the main attention to
multicolor QCD, matrix models,
loop equations, reduced models, 2D quantum gravity, free random variables,
noncommutative theories, AdS/CFT correspondence.
These issues attracted the most 
interest of the community over the last thirty years.

I apology that some of the important results, in particular
on QCD phenomenology and supersymmetric gauge theories, are not
mentioned in this short talk. Correspondingly, the list of references
is far from being complete.
An extended description of the subject of this talk
as well as more references
can be found in my recent book~\cite{Mak02}.

\section*{1974:~~~Multicolor QCD}

The effective coupling constant
in quantum chromodynamics (QCD) becomes large at
large distances where the perturbation theory is not applicable.
The idea of 't~Hooft~\cite{Hoo74} was 
to consider the dimensionality of the gauge
group $SU(\N)$ as a parameter
and to  perform an expansion in $1/\N$,
the inverse number of colors.
The motivation was an expansion in the inverse number of 
field components $N$ in statistical mechanics,
where only bubble graphs of the type  depicted in Fig.~\ref{fi:bubble} 
survive at large $N$.

The expansion of QCD in $1/\N$ (known as 
the $1/\N$-expansion) rearranges
diagrams of perturbation theory according to their
topology. Only planar diagrams of the type depicted in Fig.~\ref{fi:dual}
survive the large-$\N$ limit,
while the expansion in $1/\N$
plays the role of a topological expansion.
In the 't~Hooft limit when $\lambda=g^2 \N$ is kept fixed as $\N\ra\infty$,
a generic (properly normalized) diagram of genus $h$ with $L$
quark loops and $B$ external boundaries behaves as 
\be
\hbox{generic graph}~\sim~\left(\frac1{\N}\right)^{2h+L+2(B-1)} 
\label{ordergenericee}
\ee 
independently of the order of the diagram in the coupling constant.

This is similar to an expansion in the string 
coupling constant in dual-resonance models 
of the strong interaction, that also has a topological character
and the
phenomenological consequences of which agree with experiment. The accuracy of
the leading-order term, which is often called 
multicolor QCD or large-$\N$
QCD, is expected to be of the order of the ratios of meson widths to their
masses, \ie about 10--15\%. 

The simplification of QCD in the large-$\N$ limit arises from the fact
that the number of planar graphs grows with the number of vertices 
only exponentially rather than factorially as do the total number of graphs.
The number of graphs of genus $h$ with $n_0$ vertices grows at
large $n_0$ as
\be
\#_h(n_0)\approx  \e^{\Lambda_c n_0}(n_0)^{-b_h} \,,
\label{nographs}
\ee
where $\Lambda_c$ is a constant, so the dependence on genus resides
only in the index $b_h$ in the pre-exponential. 

While QCD is simplified in the large-$\N$ limit, it is still not
yet solved (except in $d=2$ dimensions~\cite{Hoo74b}). 
Generically, it is a problem of infinite matrices, rather than
of infinite vectors as in the theory of second-order phase
transitions in statistical mechanics.
\begin{figure}[tb]
\unitlength=1mm
\linethickness{0.6pt}
\begin{picture}(100.00,22.00)(-5,0)
\thicklines
\put(7.00,2.00){\vector(1,1){5.00}}
\put(11.00,6.00){\line(1,1){4.00}}
\put(97.00,10.00){\vector(1,1){5.00}}
\put(101.00,14.00){\line(1,1){4.00}}
\put(15.00,10.00){\vector(-1,1){5.00}}
\put(11.00,14.00){\line(-1,1){4.00}}
\put(105.00,2.00){\vector(-1,1){5.00}}
\put(101.00,6.00){\line(-1,1){4.00}}
\put(22.00,10.00){\circle{10.00}}
\put(34.00,10.00){\circle{10.00}}
\put(46.00,10.00){\circle{10.00}}
\put(56.00,10.00){\makebox(0,0)[cc]{$\cdots$}}
\put(66.00,10.00){\circle{10.00}}
\put(78.00,10.00){\circle{10.00}}
\put(90.00,10.00){\circle{10.00}}
\end{picture}
\caption[Bubble graph]%
{Bubble graph which survives the large-$N$ limit of  
$N$-component vector models in statistical mechanics.
     }
   \label{fi:bubble}
   \end{figure}
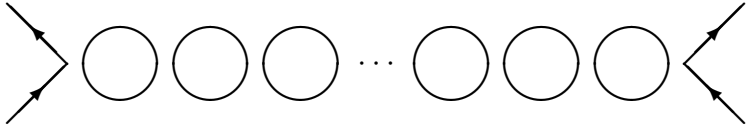
\begin{figure}[tbp]
\vspace*{3mm} \hspace*{1.8cm}
\input{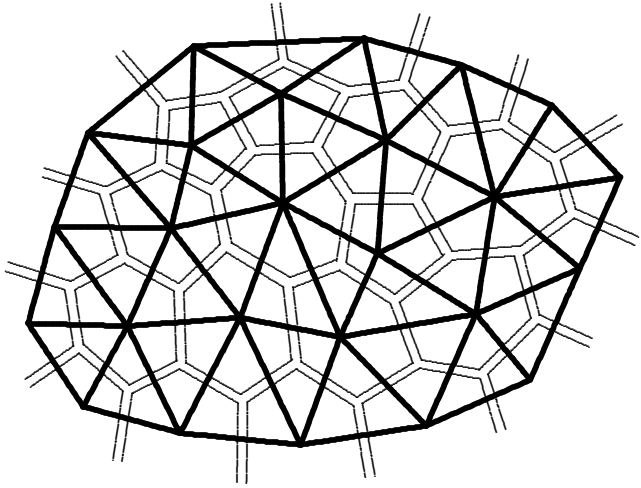} 
\vspace*{5mm}
\caption[Planar graph]%
{Planar diagram (depicted by double lines) which survives the large-$\N$ 
limit of $\N\times \N$ matrix models with cubic interaction.
Its dual graph (depicted by bold lines) is constructed from equilateral 
triangles.}
   \label{fi:dual} 
\end{figure} 
Since the correlators of gauge-invariant operators factorize in the large-$\N$
limit, it looks like the leading-order term of a 
``semiclassical'' WKB-expansion in $1/\N$. This fact was linked to the 
possible existence of a master field~\cite{Wit79} describing the
``classical'' $\N=\infty$ limit.

\section*{1978:~~~One-matrix model}

Matrix models 
first appeared in statistical mechanics and nuclear 
physics and turned out to be very useful in 
the analysis of various physical systems where 
the energy levels of a complicated
Hamiltonian can be approximated by the distribution of eigenvalues
of a random matrix. The statistical averaging is then replaced
by averaging over an appropriate ensemble of
random matrices.
 
Matrix models possess some features of multicolor QCD  
 but are simpler and can often be solved
as $\N\ra\infty$ (\ie in the planar limit) using the methods proposed
for multicolor QCD. For the simplest case of the Hermitian
one-matrix model, which is related to the problem of enumeration of graphs,
an explicit solution at large $\N$ was
first obtained by Br\'ezin, Itzykson, Parisi and Zuber~\cite{BIPZ}.
It inspired a lot of activity on this subject, in particular 
the methods to construct
the genus expansion in $1/\N$ were developed.

The Hermitian one-matrix model is defined by the partition function
\bea
Z_{\rm 1h}&=&\int \d\varphi \e^{-\N\tr{}V(\varphi)},
\label{Z1h}
\eea
where $\d\varphi$
is the measure for integrating over Hermitian $\N\times\N$ matrices.
A very important property of the model is that \/$\tr{}V(\varphi)$\/
depends only on the eigenvalues of the matrix $\varphi$.
Similarly, representing $\varphi$ in a canonical form 
$\varphi= U P\, U^\dagger$ 
with unitary $\N\times\N$ matrix $U$ and diagonal
$ P=\hbox {diag}\,\left\{p_1,\ldots,p_\N\right\}$ ,
the measure $\d \varphi$
can be written in a standard Weyl form
\be
\d \varphi=\d V \, \prod_{i=1}^\N \d p_i\, \Delta^2\!\left(P\right),
\label{Weylform}
\ee
where 
$\Delta(P)~=~\prod_{i<j} \left(p_i-p_j \right)$ 
is the Vandermonde determinant.
The contribution from angular degrees of freedom residing in $U$ 
factorizes, so the partition function~\rf{Z1h} is expressed
via $\N$ degrees of freedom.
The integral can therefore be calculated 
as $\N\ra\infty$ using the saddle-point method. 

The saddle-point equation simplifies  
for the (normalized) spectral density 
$
\rho(p)
$
which describes the distribution of eigenvalues of the matrix $\varphi$
and becomes a continuous nonnegative function of $p$ as $\N\ra\infty$.
Then the integral is dominated
by a saddle-point configuration which obeys the equation~\cite{BIPZ}
\be
V^\prime(p)=2\pint \d \lambda \frac{\rho(\lambda)}{p-\lambda} 
~~~~~~~~~~\fbox{$p\in$~support of~$\rho$}~,
\label{spe1h}
\ee
where the RHS involves the principal part of the integral.
Equation \rf{spe1h} holds only when $p$ belongs to the support of 
$\rho$.

For a general potential $V(p)$, the simplest solution is when $\rho(p)$
has support on a single interval $[a,b]$. This
 looks similar to Wigner's semicircle law for the Gaussian
case which is perturbed by the interactions.
Such a spectral density equals
\be
\rho(p)=\frac{M(p)}{2\pi}\sqrt{\left(p-a\right)\left(b-p\right)},
\label{generalrho}
\ee
where $a$ and $b$ are the ends of the support and $M(p)$ is a polynomial
of degree $K$$-$$2$ if $V(p)$ is a polynomial of degree $K$.
This solution was first obtained in~\cite{BIPZ} for cubic
and quartic potentials.

The one-cut solution~\rf{generalrho} is acceptable if $M(p)$ 
is not negative in the interval $[a,b]$ which always happens
for small values of the interaction
couplings $g_3,g_4,$ {\it etc}. With increasing couplings, a third-order
phase transition 
may occur after which a more complicated multicut solution is
realized.

\section*{1979:~~~Loop equations}

The loop-space approach in QCD was motivated by Wilson's lattice formulation
of non-Abelian gauge theories. It is based on the fact that all observables 
can be expressed at large $\N$ via quantum averages of the
trace of a non-Abelian phase factor (= the Wilson loop) 
\be
W( C ) =
\LA \frac{1}{\N} \tr {{\bbox P}}\e^{\i g \oint_C \d x^\mu A_\mu( x )}\RA.
\label{closedWloop}
\ee
Remarkably, in the large-$\N$ limit this $W(C)$ satisfies a closed
equation derived in 1979 by Migdal and me~\cite{MM79} and
known as the loop equation.

The simplest form of the loop equation I know uses
the functional Laplacian $\Delta$ which is defined as the 
$\epsilon\ra0$ limit of the second-order variational operator
\be
\Delta^{(\epsilon)} =
\int\nolimits_0^1 \d\sigma_1 \int\nolimits_0^1 \d\sigma_2\,  
\e^{-{\big|\int_{\sigma_1}^{\sigma_2}}
\d \sigma \sqrt{\dot x^2(\sigma)}\big|/{\epsilon}}
 {\delta \over \delta x_{\mu }(\sigma_1 )} {\delta \over \delta x_{\mu 
}(\sigma_2 )} \,.
\label{2.13}
\ee
The loop equation then reads as
\be
\Delta W(C)= \lambda \oint\nolimits_C \d x_\mu \pintxx \d y_\mu\,
\delta^{(d)}(x-y) W(C_{xy})W(C_{yx})\,,
\label{LE}
\ee
where $C_{xy}$ and $C_{yx}$ are the parts of the contour $C$ from
$x$ to $y$ and $y$ to $x$, respectively.

The operator $\Delta ^{(\epsilon)}$ defined by \eq{2.13}
can be inverted so that the loop equation~\rf{LE}
with the proper choice of boundary conditions can be transformed to 
the form
\be
W [x] = 1 - 
{1\over 2}\int\nolimits_0^\infty \d A \left\{\LA J [x+\sqrt{A}\xi ] 
\RA_{\xi}^{(\epsilon)}
- \LA J [\sqrt{A}\xi ] \RA_{\xi }^{(\epsilon)} \right\}\, ,
\label{4}
\ee
where $J[x]$ stands for the functional on the RHS of \eq{LE}.
In \eq{4} the average over the loops $\xi(\sigma)$ is given by the path 
integral
\be
\Big\langle F [\xi] \Big\rangle_\xi^{(\epsilon)} = 
\frac{\int_{\xi(0)=\xi(1)} D\xi \e^{-S} F [\xi]}
{\int_{\xi(0)=\xi(1)} D\xi \e^{-S}}
\label{5}
\ee
with the reparametrization-invariant local action  
\be 
S = {1\over 4}\int_0^1 \d\sigma  \left\{ 
\frac{\epsilon}{\sqrt{\dot x^2(\sigma)}}\, \dot\xi^2 (\sigma )  +
\frac{\sqrt{\dot x^2(\sigma)}}{\epsilon}\, \xi^2 (\sigma )\right\} 
\label{6} 
\ee
which is of the type of that for a harmonic oscillator.

The integral form~\rf{4} of the loop equation is most convenient
for an iterative solution in $\lambda$, which reproduces
the diagrams of perturbation theory as $\epsilon\ra0$.
The diagram with a three-gluon vertex appears thereby as a result
of doing an uncertainty of the type $\epsilon \times 1/\epsilon$.

As far as nonperturbative solutions of the loop equation are concerned,
it was shown that the area law is a self-consistent solution for 
asymptotically large smooth contours. Although this is consistent
with the string representation, it was also shown that the free 
bosonic Nambu--Goto string
is {\em not}\/ a solution for intermediate loops.
A formal solution for all loops was found~\cite{Mig81}
in the form of an elf-string with
 two-dimensional
elementary fermions living in the surface. They 
 were introduced to provide a factorization. 
For large loops the internal fermionic structure becomes
frozen, so the area law is recovered.
However, it is unclear whether or not the elf-string 
is practically useful for studies of multicolor QCD, since the methods
of dealing with the string theory in four dimensions have not
yet been developed.

\section*{1982:~~~Reduced models}

The large-$\N$ reduction 
was discovered in 1982 by Eguchi and
Kawai~\cite{EK82} and stated that the $SU(\N)$ 
gauge theory on a $d$-dimensional space-time is equivalent at $\N=\infty$ 
to the one at a single point. 
The continuum reduced action%
\footnote{Here the coefficient $(2\pi/\Lambda)^d$ represents a
``unit volume'' for a box.}
\be
S_{\rm EK}[{\bbox A}] =  - \left(\frac{2\pi}{\Lambda}\right)^{d}
\frac{1}{4g^2} \tr [{\bbox A}_\mu,{\bbox A}_\nu]^2\,,
\label{defSEK}
\ee
can be viewed as obtained from the usual one  
by substituting the covariant derivative according to
\be
\i \partial_\mu +g A_\mu(x) \stackrel{{\rm red.}}{\rightarrow}
D^\dagger(x) \, {\bbox A}_\mu\, D(x) \,,
\label{Asubstitution}
\ee
where the matrix ${\bbox A_\mu}$ is space-independent. 
The Eguchi--Kawai (EK) model was found
as a solution of the (lattice) loop equation.

The equivalence of the EK model
and the usual theory was based on an extra $R^d$ 
symmetry of the reduced action~\rf{defSEK}, which should not be broken
spontaneously.
Soon after the EK model was proposed,
it was recognized that for $d>2$ a phase transition occurs in
the EK model with decreasing coupling constant and the 
$R^d$ symmetry is, in fact, broken at weak couplings. 
The quenching prescription~\cite{BHN82},
when the eigenvalues of the (infinite) Hermitian matrix ${\bbox A}_\mu$
are quenched, was proposed to cure the construction. The quenched EK model
results in a reduced model which recovers multicolor
QCD both on the lattice and in the continuum, while it makes sense only
for planar diagrams.

An elegant alternative to the quenched EK model,
which also preserves the $R^d$ symmetry and describes multicolor QCD, 
was proposed in 1983 by Gonz\'alez-Arroyo and Okawa~\cite{GO83} on
the basis of a twisting reduction prescription. 
The corresponding lattice version 
of the twisted Eguchi--Kawai model (TEK) lives on a unit hypercube with 
twisted boundary conditions. The continuum version of TEK~\cite{GAK83}
is described by the action
\be
S_{\rm TEK}[{\bbox A}] =  - \left(\frac{2\pi}{\Lambda}\right)^{d}
\frac{1}{4g^2} \tr \left([{\bbox A}_\mu,{\bbox A}_\nu] + 
\i B_{\mu\nu} {\bbox 1} \right)^2,
\label{defTEK}
\ee 
where $B_{\mu\nu}$ is an antisymmetric constant tensor. 
It cannot be omitted in \eq{defTEK} because ${\bbox A_\mu}$
are infinite matrices (= Hermitian operators) for which
$\tr {} \left[{\bbox A}_\mu, {\bbox A}_\nu \right]\neq0$.
Although the actions~\rf{defSEK} and \rf{defTEK} look similar,
the difference between EK and TEK resides in the vacuum states
which are determined by the equation
\be
\left[{\bbox A}_\mu, {\bbox A}_\nu \right] = -\i B_{\mu\nu}{\bbox 1} 
\label{HCR}
\ee
and drastically differ from those at $B_{\mu\nu}=0$
given by diagonal (commutative) ${\bbox A}_\mu$.

The TEK model reveals interesting mathematical structures
associated with representations of 
the Heisenberg commutation relation~\rf{HCR}
(in the continuum) or its finite-dimensional approximation by
unitary matrices (on the lattice). In contrast to the quenched EK
models which describe only planar graphs, the TEK models 
make sense order by order in $1/\N$ and even at finite $\N$, 
when they are associated with gauge theories on a noncommutative 
lattice described below. 

While the reduced models look like a great simplification, since
the space-time is reduced to a point, they still involve an integration
over $d$ infinite matrices which is a continual path integral.
For some years it was not clear whether or not this is a real simplification
of the original theory that can make it solvable, so the point of view
on the reduced models 
was that they are just an elegant representation at large $\N$.

\section*{1985 \& 1990:~~~2D quantum gravity}

Matrix models are generically associated~\cite{ADF85}
with discretization of random surfaces.
The simplest Hermitian one-matrix model corresponds to a zero-dimensional
embedding space, \ie to two-dimensional Euclidean quantum gravity 
described by the partition function 
\be
Z_{2{\rm DG}} =
\int \D g \e^{-\int \d^2x \sqrt{g} \Lambda + {2(1-h)}/{\G} } \,.
\label{2Dgravity}
\ee
Here $\Lambda$ denotes the cosmological constant,
while the coupling $\G$ weights topologies of the 2D world. 
The path integral in \eq{2Dgravity} is over all metrics $g_{\mu\nu}(x)$.

The idea of dynamical triangulation of random surfaces  
is to approximate a
surface by a set of equilateral triangles. The coordination number (the number
of triangles meeting at a vertex) does not necessarily equal six, which 
represents internal curvature of the surface.
The partition function \rf{2Dgravity} is then approximated by
\bea
Z_{\rm DT} &=& \sum_h
\e^{ 2 (1-h)/\G} \sum_{T_h} \e^{-\Lambda n_{\rm t}} \,,
\label{Zdt}
\eea
where the sum over triangles is split into the sum over genus
$h$ and the sum over all possible triangulations  $T_h$ at fixed $h$.
In~\rf{Zdt}  $n_{\rm t}$
denotes the number of triangles which is not fixed at the outset,
but rather is a dynamical variable. 

The partition function~\rf{Zdt} can be represented as a 
matrix model. A graph dual to a generic set of equilateral
triangles coincides with a graph in the 
Hermitian one-matrix model with a cubic interaction as 
is depicted in Fig.~\ref{fi:dual}.
The precise statement is that $Z_{\rm DT}$ equals 
the (logarithm of the) partition function~\rf{Z1h}
with $\N=\exp{(1/\G)}$ and the cubic
coupling constant $g_3=\exp{(-\Lambda)}$. This can be easily shown by
comparing the graphs. The logarithm is needed to pick up connected
graphs in the matrix model.
Analogously, the interaction $\tr{}\varphi^k$ in the matrix model
is associated with discretization of 
random surfaces 
by regular $k$-gons, the area of which is $k$$-$$2$ times the area of the 
equilateral triangle.

Continuum limits of the Hermitian one-matrix model are reached
at the points of phase transitions. 
While no phase transition
is possible at finite $\N$ since the system has a finite number of
degrees of freedom, it may occur as $\N\ra\infty$
which plays the role of a statistical limit. 
This third-order phase transition is of the type discovered by
Gross and Witten~\cite{GW80} for lattice QCD in $d=2$
which reduces to a unitary one-matrix model. 
It is associated with divergence of the sum over graphs at each fixed
genus rather than with divergence of the sum over genera.
The contribution of a graph with $n_0$ trivalent vertices is
$\sim (-g_3)^{n_0}$ but an entropy (= the 
number) of such graphs at fixed genus is given by \eq{nographs}, 
so the sum can diverge at a certain critical value  
$g_3= \exp(-\Lambda_c)$.
This critical behavior emerges when 
one or more roots of $M(p)$  in the 
spectral density~\rf{generalrho}
approaches the end points $a$ or $b$.


This continuum limit is good for planar diagrams or genus zero, 
while higher genera are still suppressed as $\N^{-2h}$.
A description of the higher genera by the matrix models became
possible after
the ``October breakthrough'' of 1990~\cite{BK90}. It was based on the fact
that the effective parameter of the genus expansion near the critical point is 
\be
{\cal G} =\frac{1}{\N^2(\Lambda-\Lambda_c)^ {5/2}}
\ee
and can be made finite if $ (\Lambda-\Lambda_c) \sim \N^{-4/5} $ as
$\N\ra\infty$.  
This special limit, when the couplings  reach critical
values in a $\N$-dependent way simultaneously with 
$\N\ra\infty$, is called the double scaling
limit. 

The double scaling limit of the Hermite an one-matrix
model gave the genus expansion of 2D quantum gravity~\cite{BK90}.
An extension of this construction to multimatrix models made
it possible to describe  2D quantum gravity interacting with matter.
However, nobody managed so far to extend these results to higher dimensions
(beyond the so-called $d=1$ barrier).

\section*{1995:~~~Free random variables}

Since the paper~\cite{Hoo74} there were several attempts to formulate what
I would call the planar quantum field theory, \ie to describe the planar
limit in a field-theoretical language. Because external legs of a planar
graph cannot be interchanged (except in a cyclic order), this leads us
to the concept of noncommutative variables. 

The planar quantum field theory possesses a number of unusual properties.
In particular,
the usual exponential relation between the generating 
functionals $W$ and $Z$ 
for connected graphs and all graphs
does not hold for the planar graphs.
The reason for this is that exponentiation of a connected planar diagram 
can give disconnected nonplanar diagrams.
The required relation for planar graphs
can be constructed~\cite{Cvi81} by means of introducing 
noncommutative sources $\j_\mu(x)$. 
``Noncommutative'' means that there is no way to 
transform $\j_{\mu}(x)\,\j_{\nu}(y)$ into
$\j_{\nu}(y)\,\j_{\mu}(x)$. 

The conjugate variable obeys 
\be
\frac {\delta}{\delta \j_\mu(x)} \,\j_\nu(y) 
= \delta_{\mu\nu} \delta^{(d)}(x-y) \,.
\label{Cuntz}
\ee
There would be a commutator in this formula for Bosons or 
an anticommutator for Fermions, but it is just like that
in the planar limit!
It is called the Cuntz algebra and it results in 
nether Bose nor Fermi but Boltzmann statistics.

The planar contribution to the Green functions 
and their connected counterparts can be obtained, respectively, from the
generating functionals 
$Z[\j ]$ and $W[\j ]$ by
 applying the noncommutative derivative according to \eq{Cuntz} which
picks up only the leftmost variable.
The usual exponential relation is superseded by
\be
Z[\j]=W\!\left[\j Z\!\left[\j \right]\right],
\label{plconndisconn}
\ee
while the cyclic symmetry gives
$
W\!\left[\j Z\!\left[\j \right]\right]=
W\!\left[ Z\!\left[\j \right]\j\right]
$.
In other words, given $W[\j]$, one should construct
a function $\J_\mu[\j]$ which is inverse to
\be
\j_\mu(x)=\J_\mu\!\left(x\right) \frac{1}{W[\J]}\,,
\label{invfunct}
\ee
after which \eq{plconndisconn} gives
$Z[\j ]=W[\J ]$.

For the one-matrix case when there is only one source (that commutes
with itself), Wigner's semicircle law as well as some results of
Ref.~\cite{BIPZ} are reproduced by this technique.
For the $d$-dimensional Gaussian case when $W[\j]$ is quadratic in $\j$,
$Z[\j]$ can be expressed via a continued fraction. 
However, in general, mathematical methods for dealing with functions 
of this kind of noncommutative variables are not developed.

Nevertheless, there is an example of two-dimensional Yang--Mills
theory, whose solution at $\N=\infty$ was described
by Singer~\cite{Dou95}  
using an adequate mathematical language of
free random variables introduced in 
noncommutative probability theory by Voiculescu.
Such a master field quantifies earlier ideas about 
the one-matrix model~\cite{HH} and a stochastic master field~\cite{GH83}.
In some cases, an explicit solution of a $d$-dimensional planar
quantum field theory
can be easily obtained from its $d=0$ or $d=1$ counterparts, using
either additivity or multiplicativity of the free random 
variables~\cite{Dou95}.

\section*{1998:~~~Noncommutative theories}

The recent revival of the reduced models 
has arisen from the M(atrix) formulation~\cite{BFSS96} of
M-theory combining all types of superstring theories. 
The novel point of view on the reduced models, discovered in 1998
by Connes, Douglas and Schwarz~\cite{CDS98}, 
was their equivalence to gauge theories on noncommutative space,
whose coordinates do not commute and obey the commutation relation
\be
\left[{\bbox x}_\mu,{\bbox x}_\nu \right]=\i \,\theta_{\mu\nu} {\bbox 1}\,.
\label{xcommutator}
\ee
The multiplication of matrices (operators) can be represented 
in the coordinate space, introducing a noncommutative product
of functions
\be
{\bbox f} \cdot  {\bbox g} \Rightarrow  f (x) \ast g (x)\defeq
f  (x)\,\exp\left(\frac {\i}{2} \,                                        
\overleftarrow{\partial}_{\!\mu} \theta_{\mu\nu}  
{\partial_\nu}\right)\, g (x) \,,
\label{starproduct}
\ee
where $\overleftarrow{\partial}_{\!\mu}$ acts on $f  (x)$
and $
{\partial_\nu}$ acts on $g (x)$. 
This star-product is noncommutative but associative
similarly to the product of matrices (operators).

The action~\rf{defTEK} can then be rewritten as the action of
the noncommutative $U_\theta(1)$ gauge theory:%
\footnote{The tensor $\theta_{\mu\nu}$ is inverse to $B_{\mu\nu}$
entering \eq{defTEK}:
$\theta_{\mu\nu}= B_{\mu\nu}^{-1}$.}
\be
S[\A]= \frac{1}{4\lambda}\int \d^d x\;{\cal F}^2 ,
\label{NCaction}
\ee
where $\lambda=g^2\N$ coincides with the 't~Hooft coupling of the TEK model.
The gauge field $\A_\mu(x)$,
${\bbox A}_\mu\Rightarrow (\i \partial_\mu+\A_\mu)$,
is no longer matrix-valued but rather 
noncommutativity of matrices in the reduced model 
is transformed into noncommutativity of coordinates
in the noncommutative gauge theory.
In \eq{NCaction} ${\cal F}$ denotes the noncommutative field 
strength
\be
{\cal F}_{\mu\nu}=\partial_{\mu} \A_\nu -\partial_{\nu} \A_\mu -
\i \left(\A_\mu \ast \A_\nu- \A_\nu \ast \A_\mu  \right).
\label{defNCF}
\ee  
Note that cubic and quartic interactions
of $\A_\mu$ enter the action~\rf{NCaction} quite similarly to 
Yang--Mills theory! 

The diagrams of the
perturbation-theory expansion of the noncommutative theories 
look similar to those in Yang--Mills theory. 
Planar diagrams do not depend on the parameter of noncommutativity 
$\theta$ at all, in analogy with the TEK models~\cite{GO83}, and
are the same as planar diagrams in ordinary Yang--Mills theory.
For $d>2$ the contribution of a nonplanar diagram of genus $h$ is
suppressed at large $\theta$ as~\cite{Fil96} 
$\left( \det \theta_{\mu\nu} \right)^{-h}$,
so only the planar diagrams survive as $\theta\ra\infty$.
For this reason it is often said that the noncommutative $U_\infty(1)$
gauge theory is a master field of multicolor QCD.

Remarkably, the analogy between the TEK models and 
the noncommutative theories can be pursued beyond the 
$\N=\infty$ or $\theta=\infty$ limits. 
The genus expansion of the noncommutative $U_\theta(1)$ gauge theory 
can be reproduced by the TEK models in a certain double scaling 
limit~\cite{AIIKKT}. Moreover, the TEK models at finite $\N$
are mapped~\cite{ANMS99} onto noncommutative theories on
a finite periodic lattice (= a discrete torus).
These results are an extension of the fact that
noncommutative gauge theories on a torus are equivalent 
at special values of $\theta_{\mu\nu}$ to ordinary
Yang--Mills theories on a smaller torus with twisted boundary
conditions representing the non-Abelian 't~Hooft flux.

\section*{1998:~~~AdS/CFT correspondence}

A great support of the long-standing belief in the
string/gauge correspondence has come recently from 
${\cal N}=4$ supersymmetric Yang--Mills theories (SYM). 
As was conjectured by Maldacena in 1998~\cite{Mal98a}, the
${\cal N}=4$ SYM is equivalent to 
a IIB superstring in the anti-de Sitter background $AdS_5 \times S^5$.
One of the motivations for this AdS/CFT correspondence
was the underlying conformal symmetry of both theories. 
The radius $R$ of anti-de Sitter space on the superstring side
is related to 
the 't~Hooft coupling $\lambda$ on the SYM side by
\be
\frac{R^2}{\alpha^\prime}=\sqrt{\lambda}\,,
\label{lvsR}
\ee
so the strong-coupling limit of SYM is described by supergravity in 
anti-de Sitter space $AdS_5 \times S^5$.

Among the most interesting predictions of the AdS/CFT correspondence
for the strong-coupling limit of SYM, I shall mention
the calculation~\cite{GKP98}
of the anomalous dimensions of certain operators 
 and that~\cite{Mal98b} of the 
Euclidean-space rectangular Wilson loop determining the interaction
potential. 
The former is given by the spectrum
of excitations in $AdS$ space, while the latter is given by
the minimal surface formed by the worldsheet of an open string whose ends lie
at the loop in the boundary of $AdS_5 \times S^5$.
The computation of the Wilson loop in the supergravity approximation
was also performed for circular loop, which case
 has then been exactly calculated~\cite{ESZ00}
in SYM to all orders in  $\lambda$.   
The result provided not only a beautiful test of the AdS/CFT correspondence
at large $\lambda$ but also a challenging prediction for IIB superstring
in the $AdS_5 \times S^5$ background.

Yet another remarkable test of the string/gauge correspondence, 
which goes beyond the supergravity approximation, concerns~\cite{BMN02} a
certain class of operators in SYM, whose anomalous dimensions can be 
exactly computed as a function of $\lambda$ both in string theory
and under some mild assumptions in SYM.
The exact computation in string theory is possible because 
the anomalous dimensions of these
BMN operators correspond to the spectrum of states
with large angular momentum associated with rotation of an infinitely 
short closed string around the equator of $S^5$.

Rotating similarly a long closed folded string in $AdS_5$, a very
interesting prediction concerning the strong-coupling limit
of the anomalous dimensions of twist (= bare dimension minus Lorentz spin $n$)
two operators has been obtained recently in Ref.~\cite{GKP02}:
\be
\Delta -n = f(\lambda) \ln {n} 
\label{GKP}
\ee
for large $n$,
where
\be
f(\lambda)=\frac{\sqrt{\lambda}}{\pi}
+{\cal O}\left((\sqrt{\lambda})^0\right)
\label{GKP1}
\ee
for large $\lambda$. It has then been shown~\cite{Kru02} how 
this result can be reproduced via a minimal surface
of an open string spanned in the boundary by the loop with a cusp.

Equations~\rf{GKP} and \rf{GKP1} 
were derived~\cite{GKP02} ignoring the $S^5$ part of $AdS_5 \times S^5$,
which is responsible for supersymmetry, and possess the features expected
for the anomalous dimension in ordinary (nonsupersymmetric) 
Yang--Mills theory.
There are arguments for this result to be valid in ordinary Yang--Mills theory
as well. This would lead us to very interesting predictions 
for the strong-coupling limit of QCD! 
A challenging problem is whether it is possible
 to obtain \eq{GKP1} within Yang--Mills
theory, verifying thereby the string/gauge correspondence.



\end{document}